\documentstyle{article}

\begin{document}

\begin{center}
{\huge\bf On The Equivalence Of Four Dimensional And Two  
dimensional Field Theories}
\end{center}

\vspace{1cm}
\begin{center}
{\large\bf
F.GHABOUSSI}\\
\end{center}

\begin{center}
\begin{minipage}{8cm}
Department of Physics, University of Konstanz\\
P.O. Box 5560, D 78434 Konstanz, Germany\\
E-mail: ghabousi@kaluza.physik.uni-konstanz.de
\end{minipage}
\end{center}

\vspace{1cm}

\begin{center}
{\large{\bf Abstract}}
\end{center}

\begin{center}
\begin{minipage}{12cm}
We investigate the dimensional, the dynamical and the topological  
structures of four dimensional Einstein and Yang-Mills theories. It  
is shown that these theories are constructed from two dimensional  
quantities, so that they possess always a distinguished two  
dimensional substructure. In this sense the four dimensional field  
theories are equivalent to related two dimensional field theories.
\end{minipage}
\end{center}

\newpage
The theories of fundamental interactions of physics are given by  
the four dimensional field theories, from which the three Yang-Mills  
theories can be considered only as partly renormalizable quantum  
field theories \cite{ein}, whereas the Einstein theory of  
gravitation is a non-renormalizable and hence non-quantizable  
theory. The mathematical frame work of these theories is based on  
the one hand on the class of "non-trivial" and "simple" forms up to  
the two form, i. e. up to the {\it curvature} two forms \cite{tto}.  
On the other hand it is based on the class of differential operators  
up to the second order operators, i. e. up to the  
Laplace/d'Alembert operators. In other words, beyond the trivial  
volume forms, all higher forms in field theories are restricted to  
be constructed by exterior products of simple one and two forms. A  
fact which is related, in view of the relations between invariant  
dimensions of cohomology elements and invariant indicies of  
differential operators, with the mentioned restriction of physical  
differential operators to the differential operators of first and  
second order. Such relations are due to the duality between the  
homology and cohomology elements in accord with de Rham theorem and  
the equality of their dimensions: The Betti numbers. In other words  
it is due to the relation between the cohomology dimensions and the  
indicies of differential operators which determine the dynamics of  
field theories \cite{top}. In view of these restrictions, the  
physical or dynamical content of four dimensional theories, which  
can be derived from an action principle, should be equivalent to the  
physical or dynamical content of related two dimensional theories  
where the two forms are the natural geometrical limit \cite{two}.  
Thus in both cases the dynamical content is restricted to be given  
by differential operators up to the second order, which act on  
quantities up to the two forms.

Now differential operators on the physically interesting compact  
manifolds have invariant indices which do not depend on the {\it  
dimension} of the underlying compact manifolds \cite{s24}: Thus the  
topological dimensions of forms, as the dynamical quantities of  
field theories, which are given by the dimensions of the harmonic  
forms or the cohomology elements, i. e. Betti numbers, also do not  
depend on the {\it dimension} of the underlying manifold ( see  
below). Furthermore such {\it numbers} are equal for Poincare dual  
elements on a manifold in view of $H^a \cong H^{n - a}$, which  
reduces such invariant {\it numbers}, in the case of four  
dimensional theory, to the invariants of up to two forms, i. e. to  
those of the related two dimensional theory. Therefore, in view of  
such a reducibility and related circumstances one can speak from the  
reduction of four dimensional field theories on compact manifolds  
$( \sim S^4)$, to related two dimensional theories on compact  
manifolds $( \sim S^2)$, with respect to such topological  
characteristics \cite{s24}. We show after dimensional, dynamical and  
topological discussions of structure of four dimensional field  
theories that, in view of the fact that the topological invariants  
of the four dimensional field theories, i. e. the dimensions of  
cohomology/homology elements of the theory or the indicies of  
differential operators which determine the dynamics of equations of  
motion of four dimensional field theory, are given by the invariants  
of the two dimensional field theory. Therefore the four dimensional  
theories can be considered to be equivalent to related two  
dimensional field theories.
Note that the same differential operators and forms determine the  
dynamical behaviour of the physical quantities of the field theories  
for example by the number of solutions of the related differential  
equations of motion, which is given by the index of the involved  
differential operator. In other words considering the topological  
strucuture of field theories disclose also the {\it invariant} local  
or dynamical structure of these theories.

From physical point of view, the above restriction of physical  
theories to quantities which are represented by forms up to the  
second order, i. e. up to the curvature form, can be understood as  
follows: Any regular physical effect is due to some force or  
potential which is related with a curvature. Thus in view of the  
fact that a true force have dimension $\displaystyle{\frac{1}{l  
\cdot t}} \sim L^{-2}$ in geometric units \cite{form}. Therefore  
such effects, i. e. all regular physicall effects, should be of such  
a two dimensional {\it origin} ($\sim L^{-2}$) in the mentioned  
sense: Since the two dimensional structure is charcterized by its  
highest and lowest dimensions: The dimension of its area: $L^2$ and  
its curvature tensor: $L^{-2}$ (see also below).

Thus we will show that from dynamical and topological point of  
views the four dimensional theories are equivalent to the related  
two dimensional theories where the action function are given by the  
topological invariant surface integrals of related curvature two  
forms. Thus one can describe the same physical fact by a four  
dimensional theory with {\it four dependent variables} from which  
only two are independent, or by a related two dimensional theory  
with the {\it two independent variables}. Note that the two  
dimensional theories have the advantage to be integrable and hence  
quantizable, without constraints and renormalization problems of  
four dimensional theories which could not be solved yet in a  
satisfactory manner. Thus the two diomensional theory of gravity can  
be considered as the only quantizable theory of gravity, since as  
it is mentioned above the four dimensional theory of gravity is  
non-quantizable.

\bigskip
There are various hints for a two dimensional foundation of  
physical theories of fundamental interactions, from the  
mathematical- and physical frame works of such theoris: Among them,  
as already mentioned, the very fundamental fact that we are given,  
not only from physical side but even from the mathematical side,  
only simple forms up to the second order. In other words the  
mathematical quantities to describe mathematical and physical  
processes are restricted to differential forms or antisymmetric  
tensors up to two forms or second rank tensors, respectively:  
$\Omega^0 \ , \Omega^1 = \Omega_i dx^i$ and $\Omega^2 = \Omega_{ij}  
dx^i \wedge dx^j$ which correspond to the scalar functions, the  
connections or momentums or potentials and the curvatures or field  
strengths or forces on manifolds or in field theories, respectively.  
Note that all forms are of invariant dimension $L^0$, since the  
invariant dimensions of their tensor components and their coframe  
basis are reciprocal, in accord with the dimensional convention:  
$(dx \sim x \sim L)$, $((dx \wedge dy) \sim L^2)$ etc.. Thus such  
dimensions are determined in accord with the coframe structure of  
the volumes invariants on orientable manifolds: $((dx^1 \wedge, ...,  
\wedge dx^n) \sim L^n)$ \cite{form}.

In this sense the {\it symmetric} tensors are considered with  
respect to the exterior algebra as zero forms or scalar functions,  
i. e. of dimension $L^0$: Since symmetric tensors possess no  
antisymmetric components and the volumes which define the  
fundamental dimensions of manifolds and theories, are defined {\it  
entirely} by the {\it antisymmertic} coframe basis. This  
consideration is also in accord with the definition of the  
fundamental symmetric metric tensor by: $ ds^2 := g_{(ij)} dx^i  
\cdot dx^j$ where $ds^2 \sim ( dx^i \cdot dx^j )$ and also the  
metric $g_{(ij)}$, all can be considered of dimension $L^0$, in view  
of their symmetric structure.

Note that in the same sense, the tensor components of differential  
operators, such as $( \partial_i \sim  
\displaystyle{\frac{\partial}{\partial x^i}} \sim  
\displaystyle{\frac{1}{x^i}} )$ and $( \epsilon_{ij} \partial_i  
\partial_j \sim \displaystyle{\frac{1}{x \wedge y}} )$ can be  
considered of dimensions $L^{-1}$ and $L^{-2}$, respectively.  
Whereas the symmetric derivatives like Laplace operator: $\Delta  =  
(d d^{\dagger} + d^{\dagger} d)$, can be considered of dimension  
$L^0$ in accord with their symmetric structures, thus their  
application on forms does not change their order. In the same sense,  
the antisymmetric tensor components of vector- and tensor fields on  
a n-dimensional manifold has dimensions $L,..., L^n$ and their  
antisymmetric frame basis are of dimensions $L^{-1}, ..., L^{-n}$.
Note further that such invariant dimensions of tensor components  
are independent of the dimension of the underlying manifold, thus  
they are in accord with the well known physical dimensions of  
related physical quantities such as (( force or curvature or field  
strength ) $\sim L^{-2}$) and (( potential or momentum or mass )  
$\sim L^{-1}$), etc. in geometric units. Thus a force is the  
gradient of a regular potential which is a component of a connection  
one form.

Recall also that in accord with the phase space philosophy the time  
is not a true phase space variable, i. e. it is not an independent  
variable beside the canonical position and momentum variables in  
phase space, but it can be introduced only as a {\it parameter} for  
this variables. Thus the "canonical conjugate variable" of time  
parameter , i. e. the Hamiltonian function, is a {\it function} of  
momentum and position, i. e. a {\it depndent} variable. This  
circumstance reduces the number of true variables of phase space of  
four dimensional theories to three.
Thus any theory like the four dimensional field theories which  
consider, in spite of this, the time as a true variable; will  
possess constraints and the time component of its field variable  
should be eliminated by a gauge fixing condition.

There is a further hint for the fundamental role played by the two  
dimensional thories in physics, which is described by the fact that  
in view of KAM and Morse-Smale stability theorems on the  
integrability of physical sytems, only systems with two {\it  
independent} degrees of freedom are stable against small  
perturbations \cite{abma}. In other words only two dimensional  
theories, i. e. those with only {\it two independent} degrees of  
freedom, are integrable and stable with respect to small  
perturbations of related systems. Note that in this sense the {\it  
small perturbation} or deviation from the two dimensional structure  
of the system can not be considered as an independent variable or  
degree of freedom \cite{abma}.

Therefore it is plausible to consider that any physical theory  
which modelizes a {\it stable} physical process, e. g. the above  
mentioned fundamental theories, should be itself an stable model and  
hence should possess such a two dimensional foundation, in accord  
with the mentioned relation between the two dimensionality and the  
stability of sytems. This will means that integrable physical  
theories should possess two dimensional substructures.

Note further that in view of the fundamentality of such a {\it  
possible} invariant equivalence between the four dimensional and two  
dimensional field theories, the essential features of this  
equivalence should be embodied in the internal invariant structure  
of four dimensional field theories. Nevertheless the dominance of  
{\it local} point of view in theoretical physics diverted the  
interest from such features of theories. Thus the main interest in  
the {\it local approach} in physics was to obtain the equations of  
motion and their solutions, where the conditions or assumptions to  
obtain the equations of motion, and the topological or global  
character of these conditions and equations were not of much  
interest.

Therefore the simplest way to consider the invariant or topological  
character of boundary conditions and assumptions on physical  
quantities under which one obtains equations of motion, is to  
consider the invariant dimensions of physical quantities and their  
mathematical representants, i. e. the tensor components of  
differential forms, in accord with the above {\it invariant}  
determinations.

\bigskip
To investigate the aspired equivalence, recall that any reasonable  
field {\it theory} is defined on a differentiable {\it manifold} of  
some dimension $n$. Thus in the same sense that such a  
$n$-dimensional {\it manifold} implies invariant or constant volume  
of dimension: $L^n$, i. e. $L^n = (constant)$ and constant tensor  
components of highest form of dimension $L^{-n}$, i. e. $L^{-n} =  
(constant)$. The $n$-dimensional {\it theory} implies also an  
invariant or constant volume of dimension: $L^n$ and an invariant  
Lagrangian density of dimension $L^{-n}$, i. e. again $L^n =  
(constant)$ and $L^{-n} = (constant)$. Note also that the product  
$L^{-n} \cdot L^n = L^0 = (constant)$ represents the well known  
action invariant of the n-dimensional theory. Therefore it is  
possible to write an $n$ dimensional action function in two  
equivalent ways: By $\int\limits_{(n D)} \Omega^0 \cdot dx^1 \wedge,  
..., \wedge dx^n$ where the Lagrangian density is a zero form, or  
by $\int\limits_{(n D)} \Omega^n _{1, ..., n} dx^1 \wedge, ...,  
\wedge dx^n$ where the Lagrangian density is an $n$ form which is  
the Hodge dual of the mentioned zero form in $n$ diemsions. This is  
in accord with the well known Hodge duality of zero forms and the  
{\it highest forms} on a manifold, from which the $L^{-n}$  
dimensional tensor components are proportional to the $L^{-n}$  
dimensional Lagrangian density of the related theory. In other words  
any $n$-dimensional manifold and theory implies two equivalent  
characteristic invariants of dimensions $L^n$ and $L^{-n}$ and one  
invariant product of them. Thus any of such characteristic  
invariants implies the other one by definition. Hence a theory which  
possess one of these characteristic invariants can be considered as  
$n$-dimensional theory or equivalent to this, thus it possess the  
other invariant also. Furthermore a "higher" dimensional theory  
which implies lower dimensional characteristic invariants, as well  
as higher dimensional invariants, should be considered as equivalent  
to the lower dimensional theory which is defined by the related  
lower dimensional invariant Lagrangian and volume: Since the higher  
dimensional invariants in this case can be constructed from the  
lower dimensionmal invariants, e. g. by their product or by  
multiplication with dimensional constants, as in the case of  
Einstein theory (see below).

Therefore a theory with an $L^2$ invariant, or equivalentely an  
$L^{- 2}$ invariant, can be considered as a two dimensional theory  
which is formulated on a two dimensional manifold, or it is  
equivalent to such a two dimensional theory, in view of the fact  
that both $L^2$ dimensional voloume element and the "dual" $L^{- 2}$  
dimensional Lagrangian density in a two dimensional theory should  
be invariants. In other words from the invariant dimensional point  
of view, a higher dimensional theory with $L^2$ and/or $L^{- 2}$  
invariants or constants, should be equivalent to a two dimensional  
theory which possess these invariants by definition. Thus one can  
formulate a theory either with {\it higher} number of dependent  
variables on a {\it higher} dimensional manifold with "conditions"  
which reduce the number of independent variables to the independent  
ones; or one can formulate an equivalent theory with these  
independent variables on the {\it lower} dimensional manifold.

Another direct fact about the $n$ dimensionality of a field theory  
is that, if its solutions, i. e. the actual field variable of theory  
possess $n$ {\it independent} components, then the theory should be  
reduciable to an $n$-dimensional theory: Since a field theory with  
$n$ {\it independent} variables, or with a field variable with only  
$n$ independent components, can be defined as a $n$-dimensional  
theory on a $n$-dimensional manifold. In other words a theory with  
only $n$ independent solutions is {\it actually} a $n$ dimensionsl  
theory. This statement is in accord with the dynamical, invariant  
dimensional and topological statements on the $n$ dimensionality of  
a theory.

Further note that the above introduced invariant dimensions of  
tensor- and coframe basis components of forms which are independent  
of the diemnsion of the underlying manifold and therefore called  
invariant dimensions, are also independent of the local  
transformations of cocordinates. The reason is that we consider  
these invariant dimensions in accord with the {\it measures} of  
invariant integrals of forms: $\int\limits_{C_a} \Omega^a \ ; \ a  
\leq n$ on the $n$-dimensional manifold. In other words the  
invariant dimension of the coframe components of forms is considered  
to be equal to the dimension of the Cartesian coframe basis of  
forms which is equal to the invariant dimension of the invariant  
integral measure. This means that we consider the invariant  
dimensions of tensor- and coframe components of {\it equivalence  
classes} of forms with respect to their local transformations. It is  
in this sense that such a dimension is independent of local  
transformations and can be considered as a globally or topologically  
invariant dimension. In other words the introduced invariant  
dimensional consideration of components of forms, as the main  
ingredients of field theories, are in accord with their integrals,  
where $\Omega^0 = \int\limits_{C_a} \Omega^a$ and $dim ( \Omega^0 )  
= L^0 = dim ( \int \Omega^a _{ij...a} dx^i \wedge dx^j \wedge  
...\wedge dx^a ) = L^{-a} \cdot L^a$.

We will show that in accord with these statements the classical  
four dimensional theories of elctrodynamics and gravitation are  
indeed equivalent to two dimensional theories. Note however that the  
phase space structure of four dimnsional and two dimensional  
theories are different and hence their quantum structures are also  
different. Thus the two dimensional theories have the advantage to  
be renormalizable, whereas the four dimensional theories have  
essential problems with constraints and renormalization \cite{ein}.

\bigskip
To begin with the discussion of topological and dynamical  
equivalence of Einstein theory of gravity with a two dimensional  
theory: Note that the four dimensional theory possess the Einstein  
{\it constant} $\kappa$, which is an invariant of dimension $L^2$,  
that is introduced in the theory as the reciprocal $L^{- 2}$  
dimensional constant. Note further that it is this invariant which  
makes possible to formulate the Einstein-Hilbert action invariant in  
four dimensions, i. e. on a four dimensional manifold with the  
$L^4$ dimensional volume, since the dimension of the Lagrangian  
density of the theory: $( \sqrt{-g} R )$  is only $L^{- 2}$: In view  
of the $L^0$ dimensionality of metric and its determinant, and the  
$L^{-2}$ dimensionality of curvature: $( R \sim  
\displaystyle{\frac{1}{r^2}} )$.

Recall that the construction of Einstein-Hilbert action bears  
already a distinguished two dimensional structure within the four  
dimensional envelope of this theory, in view of the {\it neccessity}  
of $L^{- 2}$ dimensionality of the Ricci scalar $R$ to define this  
theory in four dimension: Since as a scalar function or zero form  
$R$ should be considered of dimension $L^0 = (invariant)$ with  
respect to the exterior calculus. Therefore in view of the  
neccessary equivalence between these two dimensions of $R$, i. e.  
$((invariant) = L^0 \cong L^{-2} )$, there is a neccessary  
distinguished two dimensional structure in this theory, which is  
just defined by the equivalence relation: $((invariant) = L^0 \cong  
L^{-2} )$. Thus the Ricci tensor and Ricci scalar {\it can} be  
considered to be of dimension $L^{-2}$, also in view of their  
extraction from the $L^{-2}$ dimensional tensor components of the  
curvature form.

Note further that the theory possess the cosmological {\it  
constant} $\Lambda$, which is an invariant of dimension $L^{- 2}$.

Therefore in accord with the above discussion of characteristic  
invariants of a n-dimensional theory, the four dimensional Einstein  
theory of gravity should be equivalent to a two dimensional theory  
of gravity which is formulated on a two dimensional manifols, in  
view of the fact that Eintein theory possess with $\kappa$ and  
$\Lambda$, both $L^2$- and $L^{- 2}$-dimensional characteristic  
invariants of a two dimensional theory.

There is a further fundamental fact that implies the two  
dimensionality of Einstein theory from algebraic point of view: If  
the equations of motion of a theory with a non-Abelian curvature, e.  
g. Einstein theory with Riemann curvature, are obtained in local  
geodesic coordinate system, then this theory can be considered to be  
equivalent to a two dimensional theory. The reason is that in local  
geodesic coordinate system the non-Abelian algebra and curvature  
transit to the Abelian algebra and curvature, respectively. On the  
other hand the Abelian algebra is equivalent to the $SO(2)$ algebra  
which defines the symmetry group on a two dimensional manifold. Thus  
an Abelian curvature is equivalent to rotation on the space of  
connections or to a curvature on the related two dimensional base  
manifold; and a theory with such a curvature should be equivalent to  
a two dimensional theory.

In retrospect, the fact that the Einstein equations of motion are  
{\it obtainable} and hence they are obtained in the local geodesic  
coordinates manifests the two dimensional nature of Einstein theory,  
although there is an equivalent method to obtain them.

\bigskip
The main differential geometric reason that the Einstein equations  
and so the whole dynamical content of the four dimensional theory of  
gravity is of two dimensional nature is the existence of the second  
Bianchi identity: $d \bar{R} = 0$ where $\bar{R}$ is the curvatur  
form $\bar{R} \in \Omega^2$.
Thus, a trace of this identity results in the "contracted" Bianchi  
identity with respect to the Einstein tensor: $\partial_i (R_{ij} -  
\displaystyle{\frac{1}{2}} g_{ij} R) = 0 \ ; \  i, j = 1, ..., 4$.  
However in view of the fact that the symmetric Einstein tensor can  
be considered as a scalar function with respect to the exterior  
calculus, i. e. in view of $(R_{ij} - \displaystyle{\frac{1}{2}}  
g_{ij} R) \in \Omega^0$: Therefore in accord with relation: $( d  
\bar{R} = 0 \leftrightarrow \partial_i (R_{ij} -  
\displaystyle{\frac{1}{2}} g_{ij} R) = 0 )$, i. e. in accord with $  
( d \Omega^2 = 0 \leftrightarrow d^{\dagger} \Omega^0 = 0 )$, the  
Einstein theory which contains such an identity is a two dimensional  
theory or equivalent to that: Since only a two dimensional theory  
implies a relation: $( d \Omega^2 = 0 \leftrightarrow d^{\dagger}  
\Omega^0 = 0 )$, in accord with the Hodge duality: $\Omega^2 \cong *  
\Omega^0$ (see also below). Thus in a two dimensional theory on the  
two dimensional manifold the action function of theory: $S_{2D} =  
\int\limits_{2D} \bar{R}, S_{2D} \in \Omega^0$ and the curvature two  
form of theory $\bar{R} \in \Omega^2$ can be considered as the dual  
zero- and two forms $\Omega^2 \cong * \Omega^0$ of the two  
dimensional- theory and manifold. Recall that this fact is  
consistent with the fact that, on the one hand the Ricci tensor is  
of dimension $L^{-2}$, in view of its extraction from the tensor  
components of the $L^{-2}$ dimensional curvature form. On the other  
hand it is  of dimension $L^0$, in view of its symmetric tensor  
character which is a zero form or scalar function. Thus the  
invariant $L^{-2}$ dimension, is not missed here in view of the  
trace operation over the symmetric Riemann tensor components of  
curvature two form; but it is just rewritten in its equivalent  
dimension: $L^0$, in accord with the invariant relation $L^{-2} =  
(constant) = L^0$ on the two dimensional manifold.

Nevertheless we prove further arguments in favour of this  
statement, which concern the conditions to obtain the Einstein  
equations, and its structurte: In other words we show in the  
following that even the invariant dimensional character of  
conditions to obtain the Eintein equations, shows a distinguished  
two dimensional substructure within the assumed four dimensional  
structure, in accord with the discussed invariant structures of the  
two and four dimensional theories:

Recall that the main condition under which one obtains the Einstein  
equations from the Einstein-Hilbert action function, is the  
vanishing of the term which contains the variation of Ricci tensor:  
$\delta R_{ij}$, in the variation of action function. Such a  
condition is fulfilled, without lose of generality, by the  
assumption of $\delta R_{ij} = 0$, and hence it is equivalent to  
such an assumption. In other words one can obtain form the same  
action function the same Einstein equations by the assumption of  
$\delta R_{ij} = 0$. Considering the equivalence of variation with  
the exterior derivative, i. e. $\delta \Omega^a \cong d \Omega^a$  
\cite{nur}, the condition to obtain the Einstein equations is given  
by: $d R_{ij} = 0$. Further note that the Bianchi identity is  
equivalent to the divergencelessness of Ricci tensor: $d^{\dagger}  
R_{ij} = 0$. These results together mean that the Ricci tensor which  
obeys the Einstein equation, is a harmonic or constant function,  
since both its derivatives are zero. Thus symmetric tensors are  
considered as functions with respect to the exterior calculus and  
the Hodge-de Rham theory. Hence in view of $L^{- 2}$ dimensionality  
of such a Ricci tensor and its constancy, i. e. in view of $L^{- 2}  
\sim (constant) = L^0$, the underlying Einstein theory possess by  
such a $R_{ij}$ a constant of dimension $L^{- 2}$. Therefore in  
accord with the dimensional analyses, the Einstein theory is, in  
principle, a two dimensional theory in agreement with the above  
analysis of the Einstein constant and the cosmological constant. In  
other words the two dimensionality of Einstein theory or its  
equivalence to a two dimensional theory is embodied in the stucture  
of this theory and is incorporated in several fundamental properties  
and quantities of the theory.

Recall also that the wave equation for gravitational field which  
results from Einstein equations by linearization, i. e. $\Box g_{ij}  
= 0$, is no more than the statement of harmonicity of metric field.  
Further recall that such a metric which obeys the Einstein  
equations, possess only two components or two degrees of freedom,  
which are manifested by the two directions of polarizations of  
gravitational waves. Accordingly also the Riemann- and Ricci tensors  
which obeys the Einstein equations in form of wave equation, should  
possess only two independent components, since they are obtained  
from such a two component metric \cite{stef}. Moreover note that in  
view of the fact that the gravitational field is represented,  
equivalently, by the metric $g_{ij}$- or by the Ricci field  
$R_{ij}$, therefore the dynamical equations of gravitational field  
can be represented also by the harmonicity of Ricci field: $\Box  
R_{ij} = 0$, in accord with the above statements. Therefore also the  
actual or dynmical Ricci field which obeys the Einstein equation,  
possess only two degrees of freedom, which are represented by the  
two non-vanishing components of Ricci tensor for gravitational wave,  
that are given by the second time derivative of the two components  
of the metric \cite{stef}. Nevertheless one can prove the two  
componentness of the curvature- or Ricci tensor also directly:

Thus the ten symmetric components of Ricci tensor should obey in  
addition to the four conditions to fulfil the Bianchi identities:  
$d^{\dagger} R_{ij} = 0$,   also the four variation conditions:  
$\delta R_{ij} = d R_{ij} = 0$ to obtain the Einstein equations, as  
it is discussed above. The reason that the number of last conditions  
is four, is that the symmetric Ricci tensor should be considered as  
a scalar function with respect to the exterior differentiation $d$.  
Then in accord with this condition, the one form $d R_{ij} \in d  
\Omega^0$ on the space-time {\it four} manifold, should possess {\it  
four} vanishing components in order to fulfil the variation  
conditions. Therefore only two components of Ricci tensor which  
obeys the Einstein equations and Bianchi identities, remain  
independent. Hence, in accord with the statement on the independent  
number of dimensions of a theory and the number of its independent  
solutions, the Einstein theory of gravity is a two dimensional  
theory; or at least, it is dynamically equivalent to such a two  
dimensional theory.

Note with respect to the four last conditions that, as it was  
asserted above, the symmetric metric- and Ricci tensors are indeed  
zero forms, i. e. scalar functions, with respect to the exterior  
differential calculus: Thus the above mentioned relation:  
$d^{\dagger} R_{ij} = 0$ which are equivalent to  Bianchi identity,  
can be considered indeed as an identity, since in view of the  
absence of less than zero forms one has: $d^{\dagger} \Omega^0  
\equiv 0$ or $d^{\dagger} R_{ij} \equiv 0$. Such an accordance  
between the identity: $d^{\dagger} R_{ij} \equiv 0$ and the identity  
of divergencelessness of zero forms or scalar functions, shows the  
consistency of our considerations, in accord with our consideration  
of symmetric tensors of second rank as zero forms  with respect to  
the exterior differential calculus.

Note further, the fact of harmonicity of gravitational field and  
that it possess only two degrees of freedom, can be considered as  
the main invariant statement of Einstein equations. Nevertheless  
such a harmonicity and two dimensionality of gravitational field can  
be described also by the two dimensional gravitational curvature  
field with its two dimensional action function: $S (2D) =  
\int\limits_{(2D)} \bar{R} (2D)$ over a suitable two dimensional  
submanifold of the original four manifold, which should belong to  
the second homology class, in view of invariant theoretical  
requirements (see below). Here $\bar{R} (2D) = d \Gamma (2D) =  
R_{mn} d x^m \wedge dx^n s , \ m, n = 1, 2$ is the curvature form in  
two dimensions: Since the equations of motion which follow from  
this action by its vriation with respect to the gravitational  
connection form $\Gamma (2D)$ are: $d ^{\dagger} \bar{R} = 0$. This  
equation, together with the trivial clossedness of two forms on a  
two dimensional manifold: $d \bar{R} (2D) \cong 0$, result in the  
harmonicity of the {\it two dimensional} gravitational curvature  
field $\bar{R} (2D)$. Thus such a two dimensional result is  
equivalent to the above discussed two dimensional result for the  
four dimensional gravitational field or the Ricci tensor which obeys  
the Einstein equations. Thus in view of the fact that the Ricci  
tensor is given by the mixed trace of tangential- or algebraic  
components of curvature two form: The invariant result of Einstein  
equations, that their solutions possess only two degrees of freedom,  
should be obtainable from the equations of motion which are  
obtained from the two dimensional action function of  
Einstein-Hilbert type in two dimensions: $S_{2D} =  
\int\limits_{(2D)} \sqrt{ - g(2D)} R(2D)$.

Therefore the dynamical content of the four dimensional Einstein  
theory of gravity is equivalent to the dynamical content of the two  
dimensional theory of gravity, in accord with the above discussions.

The topological equivalence between the four dimensional Einstein  
theory of gravity and the discussed two dimensional theory is partly  
obvious from the above invariant dimensional considerations. To see  
this directly, note that on the one hand, in view of the principle  
of least action, the vanishing of variation of Einstein-Hilbert  
action function results in Einstein equations of motion which are  
equivalent to the harmonicity of Ricci tensor: ($\delta S_{(EH)} = 0  
\leftrightarrow R_{ij} \in Harm^0$), in accord with the above  
discussions of variation condition and Bianchi identities. On the  
other hand, since the Einstein-Hilbert action function is a zero  
form, therefore the vanishing of its variation or its gradient:  
$\delta S_{(EH)} = d S_{(EH)} = 0, \ S \in \Omega^0$, means that it  
is a harmonic zero form, i. e.
($\delta S_{(EH)} = 0 \leftrightarrow S_{(EH)} \in Harm^0$)  
\cite{nur}. Therefore this theory possess the relation: ($S_{(EH)}  
\in Harm^0 \leftrightarrow R_{ij} \in Harm^0$), thus both $S_{(EH)}$  
and $R_{ij}$ are zero forms with respect to the exterior calculus.
Further, in view of the fact that the Ricci tensor is a mixed  
traced tensor component of the four dimensional curvature two form:
$\bar{R} (4D) \in \Omega^2$: Therefore its harmonicity can be  
obtained from, and hence it can be considered to be equivalent to,  
the harmonicity of cuvature two form:
$( R_{ij} \in Harm^0 \cong \bar{R} \in Harm^2 )$. Therefore, in  
this theory, the harmonicity of the action zero form $S_{(EH)}$  
results in the harmonicity of the curvature two form $\bar{R} (4D)$  
and vice versa, i. e.:
$( S_{(EH)} \in Harm^0 \leftrightarrow \bar{R} (4D) \in Harm^2 )$  
or they are equivalent to each other:
$Harm_{(G)} ^0 (4D) \cong Harm_{(G)} ^2 (4D)$. Here the subscript  
$(G)$ means the gravitational field on a gravitational manifold,  
where other interactions are neglected.

Nevertheless such a relation is a typical relation of a two  
dimensional theory, i. e. on a two dimensional manifold, since only  
in a two dimensional theory or on a two dimensional manifold one has  
the very general invariant relation: $Harm^0 (2D) \cong  Harm^2  
(2D)$ in accord with Poincare duality and Hodge's theorem.

Therefore in view of the fact that the actual gravitational field  
$R_{ij}$ or $g_{ij}$ of the four dimensional theory of gravity which  
fulfil the equations of motion, express the above invariant  
relation and possess only two independent components, so the theory  
can be considered to be equivalent to a two dimensional theory of  
gravity, in accord with the above discussions also from the  
invariant theoretical point of view.

Now we investigate the dynamical, the invariant dimensional and the  
topological structures of Yang-Mills theories by the example of  
four dimensional electrodynamics which is the only stablished  
Yang-Mills theory \cite{ein}:

Here first recall that only {\it two} of six components of the  
antisymmetric field strength tensor $F_{ij}$, which obey the  
identities $d F (A) = d d A \equiv 0$, are independent components:  
Since these identities can be written in the component form by $  
\omega^i = \epsilon^{ijkl} \partial_j F_{kl} = 0$ which are  
$4$-relations between the $6$-components of $F$. Thus the actual  
number of indepndent components of the electromagnetic field is $( 6  
- 4  = 2 )$ which is in accord with the well known fact that the  
photon posses only two polarization directions or two components.  
Therfeore in view of our statement about the relation between the  
independent number of field variables of a field theory and the  
actual dimension of the theory, the four dimensional theory of  
elctrodynamics is equivalent to a two dimensional theory of  
electrodynamics.

Further recall that, as it is mentioned above, from the two groups  
of Maxwell equations: $d^{\dagger} F (4D)= 0$ and $d F (4D) \equiv  
0$ which follows from the action function: $S_{(4D)} =  
\int\limits_{(4D)} F (A (4D)) \wedge * F (A (4D))$, the second group  
are identities, in view of $F := d A$. Thus the non-trivial part of  
Maxwell equations, i. e. the first group, with its two {\it  
independent} components does not need to be derived from a four  
dimensional theory. In other words the first group of these  
equations for a two component field strength form: $F (2D)$, can be  
derived in accord with the usual variation with respect to $A (2D)$,  
from a two dimensional theory with the action function: $S_{(2D)} =  
\int\limits_{(2D)} F(A (2D))$ over a suitable two dimensional  
submanifold of the original four manifold which should belong to the  
second homology class in view of invariant theoretical requirements  
(see below). Note also that, as it is known from Maxwell theory,  
even in this four dimensional theory the actual electromagnetic  
field $F (A (4D))$ or $A (4D)$ possess only two degrees of freedom  
or two independent actual components. Thus one reduces the assumed  
four degrees of freedom to these two by the application of gauge  
conditions. Hence, in view of the fact that the two dimensional  
field $F(A (2D))$ also possess two degrees of freedom, by  
definition, therefore the two dimensional field $F(A (2D))$ with its  
two dimensional action invariant: $S = \int\limits_{(2D)} F(A  
(2D))$ can also represent the usual elctromagnetic field with its  
two degrees of freedom. Thus the non-trivial equations of motion in  
both cases are the same.

From invariant dimensional point of view the harmonicity of $F$ in  
view of Maxwell equations, i. e. the constancy of its tensorial  
components $F_{ij} (4D)$, means that $L^{-2} = (constant) = L^0$,  
since dimension of $(F_{ij})$ is $L^{-2}$. This is but, as we  
discussed for the general case of $n$-dimensional theories, a two  
dimensional relation, since only on a two dimensional manifold and  
in a two dimensional theory, the $L^2$-dimensional area and so the  
$L^{-2}$-dimensional Lagrangian density are invariants or constants.  
Thus this is exactly what one has in the two dimensional theory of  
elecrodynamics with the action function $S = \int\limits_{(2D)} F(A  
(2D))$ and the equations of motion: $( d^{\dagger} F (2D)= 0$ and $d  
F (2D) \equiv 0) \sim ( F (2D) \in Harm^2)$ .

Hence the dynamical structure of electromagnetic field and its true  
number of degrees of freedom are well represented and described by  
the two dimensional theory of electrodynamics. Therefore from the  
dynamical and invariant dimensional point of views the four  
dimensional theory of electrodynamics can be considered as  
equivalent to its two dimensional theory.

The toplogical equivalence between the four dimensional Yang-Mills  
theory and its two dimensional theory is given beyond of the  
discussed invariant dimensional considerations, by the following  
aspects of topological contents of these two theoris:

First note in this relation that invariants of a four dimensional  
manifold and hence of a four dimensional theory, are given by the  
dimensions of cohomology elements or of harmonic forms, i. e. by  
Betti numbers: $b^4 (4D) = dim H^4 (4D), b^3 (4D) = dim H^3 (4D),  
b^2 (4D) = dim H^2 (4D), b^1 (4D) = dim H^1 (4D)$ and $b^0 (4D) =  
dim H^0 (4D)$. Nevertheless in view of Poincare dualities on the  
this manifold, i. e.: $H^4 (4D) \cong H^0 (4D)$ and $H^3 (4D) \cong  
H^1 (4D)$ one has $b^4 = b^0$ and $b^3 = b^1$. Therefore all  
invariants of a four dimensional field theory can be given by only  
dimensions of cohomologies up to the second order: $H^2, H^1$ and  
$H^0$, i. e. by: $b^2 (4D), b^1 (4D)$ and $b^0 (4D)$; which are also  
the cohomologies of two manifolds or of two dimensional field  
theories: $b^2 (2D) = dim H^2 (2D), b^1 (2D) = dim H^1 (2D)$ and  
$b^0 (2D) = dim H^0 (2D)$. Thus in view of de Rham theorem one can  
choose a homology basis where: $b^2 (4D) = b^2 (2D)$, $b^1 (4D) =  
b^1 (2D)$ and $b^0 (4D) = b^0 (2D)$ \cite{top}: Since also in the  
four dimensional case the Betti numbers are given by the integrals  
of related cohomology elements over the related homology elements,  
which are two and one dimensional submanifolds of the four  
dimensional manifolds. In oder words the invariants of four  
dimensional case are given by the invariants of the two dimensional  
substructures. This fact shows on the one hand the fundamental  
relation between the four dimensional field theories and the two  
dimensional field theories as intended in this work, in view of  
their common differential structure basis up to the second order and  
the resulting common invariants. On the other hand it underlines so  
the topological background of the restriction of four dimensional  
field theories to differential forms up to the second order.

Note that in a four dimensional or two dimensional manifold without  
boundary $b^1 = b^3 = 0$ or $b^1 = 0$. Thus the desired {\it two  
dimensional} field theory can be formulated on the mentioned two  
dimensional submanifold of the four manifold of the original four  
dimensional field theory, i. e. on the mentioned second homology  
manifold, which enables also the definition of Betti number $b^2  
(2D)$ in the four dimensional case. Thus in the four and two  
dimensional cases without boundary, all invariants can be obtained  
from these two: $b^2$ and $b^0$, in view of the above discussion.

Therefore in view of the above discussed and also following  
consideration, one can stablish a closed toplogical equivalence  
between the four- and two dimensional field theories:

1. In view of the mentioned restriction of mathematical-physical  
simple forms to zero- , one- and two forms, the invariants of a four  
dimensional Yang-Mills theory result from the invariants of these  
forms, i. e. from the invariants of cohomology elements $H^0 (4D)  
\in \Omega^0 (4D), H^1 (4D) \in \Omega^1 (4D)$ and $H^2 (4D) \in  
\Omega^2 (4D)$.

2. The invariants of $H^0 (4D), H^1 (4D)$ and $H^2 (4D)$ are given  
by their Betti numbers which are defined by: $b^0 = dim H^0 = dim  
Harm^0 \ , \ b^1 = \oint\limits_{C_1} \Omega^1 (4D)$ and $b^2 =  
\int\limits_{C_2} \Omega^2 (4D)$, respectively; where the  
integrations of cohomology elements $\Omega^1 (4D)$ and $\Omega^2  
(4D)$ are considered over the related homology elements $C_1$ and  
$C_2$ which are, respectively, one- and two dimensional submanifolds  
of the four dimensional basis manifold. Thus in this sense in  
accord with de Rham duality $b^a = b_a = dim H_a$, the classes of  
homological submanifolds of the four manifold, {\it up to the two  
dimensional ones}, play the same essential role like the cohomology  
elements up to the second order in the topology and invariant  
aspects of four dimensional theories.

3. From (1.) and (2.) it follows that the invariants of a four  
dimensional Yang-Mills theory result from the mentioned Betti  
numbers as the dimensions of the harmonic- or cohomolgy- or homology  
elements on the four dimensional maniofold.

4. The invariants of the above discussed related two dimensional  
theory, result from the invariants of $H^0 (2D) \in \Omega^0 (2D),  
H^1 (2D) \in \Omega^1 (2D)$ and $H^2 (2D) \in \Omega^2 (2D)$ on the  
underlying two dimensional manifold of the theory which is a {\it  
submanifold} of the four dimensional manifold. These are given, as  
in (2.) by $b^0 = dim H^0 = dim Harm^0 \ , \ b^1 =  
\oint\limits_{C_1} \Omega^1 (2D)$ and $b^2 = \int\limits_{C_2}  
\Omega^2 (2D)$, respectively; where the integrations are considered  
over the homology elements $C_1$ and $C_2$ which are, respectively,  
one- and two dimensional submanifolds of the two dimensional basis  
manifold.

5. From (3.) and (4.), i. e. from (1.) - (4.), it follows that the  
invariants of the four dimensional Yang-Mills theory should be the  
same as the invariants of the related two dimensional theory.

6. The result in (5.) is in accord with the invariants of the four  
dimensional theory and of the related two dimensional theory which  
are deduced from their equations of motion, in view of the above  
discussed invariant dimensional considerations.

Furthermore note that considering the action function as a zero  
form which is a harmonic form in view of the principle of least  
action, if we identify as above the variation of action zero form,  
by its exterior derivative \cite{nur}. Then the resulting equations  
of motion in the four dimensional Yang-Mills theory, i. e. the  
harmonicity of the $F$ form: or $F \in Harm^2$, manifests the  
following invariant relation: That the harmonicity of the action  
zero form of the four dimensional Yang-Mills theory results in the  
harmonicity of the four dimensional Yang-Mills two form and vice  
versa. In other words in the four dimensional Yang-Mills theory  
these two harmonicities are isomorphic to each other:

\begin{equation}
Harm^0 _{YM} (4D) \cong  Harm^2 _{YM} (4D)
\end{equation}

Nevertheless such a relation manifests a two dimensional theory  
with the above  discussed action function $S_{(2D)} =  
\int\limits_{(2D)} F$, since only in a two dimensional theory or on  
a two dimensional manifold one has the very general invariant  
relation: $Harm^0 (2D) \cong  Harm^2 (2D)$, in accord with Poincare  
duality and Hodge's theorem. Thus this result is in accord with the  
above result for the Betti numbers of both theories, in view of the  
fact that $b^a = dim H^a = dim Harm^a$.

This proves again the topological equivalence of the four  
dimensional Yang-Mills theory and its two dimensional counterpart.

Therefore one can understand why one was forced to reduce the  
number of components of electromagnetic field in the four  
dimensional electrodynamics from four to the two independent ones by  
gauge conditions. The reason is that the four dimensional abelian  
Yang-Mills theory describes, as we showed, from dynamical and  
invariant theoretical point of views, nothing else than the two  
dimensional dynamical and invariant facts, in view of the only two  
independent dynamical degrees of freedom of elctromagnetic field and  
in accord with the two polarization of photons.

Furthermore with respect to both type of two dimensional theories  
on compact manifolds note that, since on a two dimensional manifold  
the general element of the second cohomology group $\Omega^2 \in  
H^2$ is given by: $\Omega^2 = d \Omega^1 \oplus Harm^2$ \cite{top}.  
Therefore Gauss's law: $\int\limits_{S^2} F = \int\limits_{S^2} d A  
= 0$ does not apply here, since: $\int\limits_{S^2} \Omega^2 \neq  
\int\limits_{S^2} d \Omega^1$. Thus the integral $\int\limits_{S^2}  
\Omega^2$ is an invariant of $S^2$ which is a multiple of the area  
of $S^2$, and is proportional to the value of the applied constant  
field strenght or the constant curvature: $Harm^2$ on $S^2$, since  
not only the area is an invariant of $S^2$, but also the $Harm^2$.

Thus we proved by various consistent methodes the equivalence of  
both types of four- and the related two dimensional theories with  
respect to their dynamical, invariant dimensional and topological  
contents.

In conclusion let us recall that the mentioned topological  
invariants, such as Betti numbers $b^2$ or $b^1$, are related with  
the main quantum numbers in Bohr-Sommerfeld qunatization and in flux  
quantization, which are integrals of closed one or two forms over  
two or one dimensional manifolds, respectively. Thus there is no  
known main {\it quantum number} which is related with higher Betti  
numbers than the second one.

\bigskip
Footnotes and references


\begin{thebibliography}{100}
\bibitem{tto}
We mean by "non-trivial" forms on a $n$-dimensional manifold, those  
beyond the $n$-dimensional volume forms: $dx^1 \wedge dx^2 \wedge,  
...,\wedge dx^n$ which is dual to the zero form $\Omega^0$. Thus we  
mean by "simple" forms, those which are not given by the exterior  
product of other forms. In this sense the "non-trivial" and "simple"  
forms are connection one forms $\Omega^1$ and curvature two forms  
$\Omega^2$. The Chern-Simons form: $\Omega^1 \wedge \Omega^2$ and  
the Yang-Mills form: $\Omega^2 \wedge * \Omega^2$ are examples of  
non-simple or product forms.
\bibitem{nur}
This is the only possibility of a non-trivial definition of  
variation for functions, since a function is, as a zero form,  
divergenceless by definition: $d^{\dagger} \Omega^0 \equiv 0$, and  
$R_{ij} \in \Omega^0$. Note that also in the phase space, the  
variation of variable functions are considered to be equal to their  
differentials, i. e. $\delta P_i = d P_i$ and $\delta Q_i = d Q_i$
\end{thebibliography}
\end{document}